\documentclass[epsf,twocolumn,preprintnumbers]{revtex4}
\usepackage{graphics}
\usepackage{graphicx}% Include figure files
\usepackage{dcolumn} % Align table columns on decimal point
\usepackage{bm}
\usepackage{color}
\usepackage{epsfig}
\usepackage{comment}
\newcommand{\Sno}{Nd$_{0.8}$Sr$_{0.2}$NiO$_2$}

\begin{document}
\title{Fluctuation-frustrated flat band instabilities in NdNiO$_2$
}
\author{Mi-Young Choi$^1$}
\author{Warren E. Pickett$^{2}$}
\email{wepickett@ucdavis.edu} 
\author{Kwan-Woo Lee$^{1,3}$}
\email{mckwan@korea.ac.kr} 
\affiliation{
 $^1$Department of Applied Physics, Graduate School, Korea University, Sejong 30019, Korea\\
 $^2$Department of Physics, University of California Davis, Davis CA 95616, USA\\
 $^3$Division of Display and Semiconductor Physics, Korea University, Sejong 30019, Korea
}
\date{\today}
%\pacs{}
\begin{abstract}
The discovery that Nd$_{1-x}$Sr$_x$NiO$_2$, with the CaCuO$_2$ infinite-layer
structure, superconducts up to 15 K around the hole-doping level
$x$=0.2 raises the crucial question 
of its fundamental electronic and magnetic processes.  
The unexplained basic feature that we address 
is that, for $x$=0 and as opposed to strongly antiferromagnetic (AFM)
CaCuO$_2$, NdNiO$_2$ with the same structure and formal 
$d^9$ configuration does not undergo AFM order. We study this issue not
in the conventional manner, as energetically unfavored or as frustrated 
magnetic order, but as an instability of the AFM phase itself. 
We are able to obtain the static AFM ordered state, but find that
a flat-band, one-dimensional-like van Hove singularity (vHs) is pinned to 
the Fermi level. This situation is unusual in a non-half-filled, effectively
two-band system. The vHs makes the AFM phase unstable to 
spin-density disproportionation, breathing and half-breathing 
lattice distortions, and (innate or parasitic) charge-density disproportionation. 
These flat-band instabilities, distant relatives of single band cuprate models, 
thereby inhibit but do not eliminate incipient AFM tendencies at low temperature. 
The primary feature is that a pair of active bands 
($d_{x^2-y^2}$, $d_{z^2})$ eliminate half-filled physics and, due to 
instabilities, preclude the AFM phase seen in CaCuO$_2$. 
This strongly AFM correlated, conducting spin-liquid phase with
strong participation of the Ni $d_{z^2}$ orbital, forms the platform for 
superconductivity in NdNiO$_2$.
\end{abstract}
\maketitle

\noindent
\section{Introduction}
A disparate variety of interacting 
models \cite{c.wu2019,leonov2020,mjhan2020,wu2020,werner2020,zhang2020,lechermann2020,zhang-jin2020,kuroki2019,gu2019,chang2019,berciu2020,zhao2019,karp2020}
based on density functional theory (DFT)
studies \cite{anisimov1999,lee2004,jiang2019,nomura2019,botana2020,choi2020,ZLiu2019,geisler2020}
have been proposed to illuminate the electronic and magnetic behavior underlying 
the discovery \cite{H.Hwang2019} and experimental study \cite{hepting2020,QLi2019,fu2020} 
of long-sought superconductivity in a layered nickelate, specifically hole-doped NdNiO$_2$ (NNO). 
With the same infinite-layer structure as CaCuO$_2$ (CCO), 
which superconducts up to 110 K when doped \cite{cco110}, NNO displays several 
differences with CCO. Undoped NNO is conducting; CCO is insulating; CCO orders 
antiferromagnetically, whereas no signature of order is seen in NNO (and its iso-electronic
but non-superconducting sister LaNiO$_2$ (LNO) \cite{crespin,hayward99,kaneko09,ikeda13}) and
there is no sign of heavy fermion screening of Ni moments. CCO and NNO have 
a common $d^9$ (formal) configuration (one hole) on the transition metal ion, however 
NNO persists as (poorly) conducting to low temperature.  Central differences include positioning of 
metal $3d$ levels relative to O $2p$ levels that are substantially different in the two 
compounds \cite{botana2020}, 
the presence of Nd $5d$ character at the Fermi level \cite{lee2004,choi2020},
and the interesting effects of Nd $4f$ character in the magnetic system \cite{choi2020}.

The electronic structures of the $d^9$ infinite layer cuprates and nickelates were compared 
some time ago, with some similarities but substantial differences being 
noted \cite{anisimov1999,lee2004} %%,jiang2019} 
and quantified more recently by Wannier function analysis \cite{nomura2019,botana2020}, and
some impact of the Nd moments has been noted \cite{choi2020}. The similarities but substantial
difference continue to dominate the discussion and inform the models cited above. No items
however seem as fundamental as the difference in ground states of the undoped materials: 
antiferromagnetic (AFM) insulator for CCO, and disordered moment conductor for NNO, the latter
being a conducting quantum paramagnetic (MQPM) phase in the Sachdev-Read 
classification \cite{sachdev}.   
Based on an interacting two-band model, Werner and Hoshino \cite{werner2020} 
discussed NNO in terms of the spin-freezing theory of unconventional 
superconductivity \cite{werner1,werner2}, and the potentially simpler case of superconductivity
at high pressure in elemental Eu with disordered $f^7$ moments \cite{stpi2019}
might share in this behavior.

Here, we focus on this absence of (AFM) ordering of Ni moments in NNO, moments whose ordering has been 
the overriding hallmark of the undoped phase in cuprates. In closely related layered 
nickelates \cite{pardo2011,botana2016,mimicking,nickelates,La438} 
({\it viz.} La$_3$Ni$_2$O$_6$, La$_4$Ni$_3$O$_8$, La$_4$Ni$_3$O$_{10}$) even with non-integral
formal Ni valences magnetic ordering and sometimes 
charge ordering occur; none has a ground state without symmetry breaking, unlike NNO.
The conventional means of addressing the appearance of magnetic order is
as an instability of the nonmagnetic metallic state, either via Stoner or RKKY instabilities
as indicated by the static susceptibility $\chi(\vec q)$ or as instability to insulating
character with interatomic exchange constants. Leonov {\it et al.} provide an example of
the conventional approach \cite{leonov2020} applied to NNO (actually LaNiO$_2$) 
and extended to the DFT+DMFT (dynamical mean field theory). 

In spite of being energetically favored in the absence of fluctuations ({\it i.e.}
in DFT calculations \cite{lee2004,botana2020,choi2020}) 
and evaluations of near neighbor exchange coupling in the 10-18 meV (120-200 K)
range \cite{leonov2020,berciu2020,ZLiu2019},
the AFM ordered state 
%% $S$=$\frac{1}{2}$ quantum fluctuations tempered by small exchange interlayer 
is not observed, suggesting it is a metastable but physically 
inaccessible phase. The AFM state of NNO is however accessible computationally, allowing 
study of its electronic structure and microscopic processes that {\it render AFM order
unstable}, thereby giving insight into the inaccessibility of the AFM phase.

We should note that the questions we address lie exactly in the broad controversy
in this field: Is this nickelate basically similar to the infinite layer cuprates,
or essentially different? The similarities have been emphasized by, for
example, Botana and Norman, who obtain significant but not defining
differences in some of the
electronic characteristics of LaNiO$_2$ and CaCuO$_2$, and also offer that
doping tends to make them even more similar\cite{botana2020}. 
Jiang and collaborators contend that
the underlying Ni ion (and its environment) is essentially different from the
cuprates, being at its core nonmagnetic, and they offer therefore that the
pairing mechanism must be basically different from the cuprates\cite{berciu2020}. 
Lechermann reaches related conclusions\cite{lechermann2020}, but there are
stimulating results on both sides of this question.
Our work expands on our earlier study\cite{lee2004} and
tends to accentuate differences from the cuprates, rather than similarities.

Our methods are described in Sec. II, followed by a brief discussion of relative
energies and magnetic moments in Sec. III. Section IV provides a description of
the (experimentally inaccessible) AFM ordered state, pointing out the effects
of electronic correlation due to the on-site repulsion $U$. The Fermi surfaces
of this AFM state, and their change with low levels of doping, are described
in Sec. V. Sections VI and VII give a description of the magnetic (and charge)
and lattice instabilities, respectively, that arise. A short wrap-up of our
work is presented in Sec. VIII.

\section{Overview of Methods} 
To treat the strong intra-atomic Coulomb interactions on Nd and Ni ions,
we adopt the generalized gradient approximation\cite{gga} 
plus fully anisotropic Coulomb $U$ (GGA+$U$) method\cite{ylvisaker2010}, 
as implemented in the accurate all-electron full-potential code {\sc wien2k} \cite{wien2k}. 
As described in our previous work\cite{choi2020}, 
this approach retains the all-electron character of Nd 
which provides intra-atomic Nd $d-f$ exchange coupling but is not the subject here. 
%The reported lattice parameters (see below) were adopted \cite{H.Hwang2019}. 
With the lattice parameters experimentally observed for the superconducting thin films of \Sno~\cite{H.Hwang2019},
a $\sqrt{2}\times\sqrt{2}$ supercell is used to investigate the AFM state,
where both Ni and Nd layers have antialigned moments.
In the supercell the frozen O full-breathing motion is investigated,
whereas the half-breathing motion is calculated in a $2\times2$ supercell (see below).
%These are depicted in Fig. 4(a).
In the infinite layer structure, the Nd ion lies above/below the NiO$_2$ square by
$c/2$=1.685 \AA, the Ni-O distance is 1.96 \AA.
Description of lattice displacements are described as they are encountered in the paper.
%We use Hund's exchange coupling $J^H_{Ni}=0.7$
%eV throughout, and quote values of $U^{d}_{Ni}$ (simply called $U$) that we explore.
%$U_f^{Nd}$= 8 eV was used throughout.

For purposes of comparison, we provide comparisons of crystal structures and ionic
sizes of NdNiO$_2$ and CaCuO$_2$. For NNO, 
$a$=3.92 \AA, $c$=3.37 \AA~\cite{H.Hwang2019}; for CCO,
$a$=3.86 \AA, $c$=3.20 \AA~\cite{Siegrist1988}, 
with cell volumes 51.8 \AA$^3$ and 47.7 \AA$^3$,
respectively, an 8\% difference. The relevant Shannon-Prewitt ionic radii are 
Nd$^{3+}$: 1.25 \AA, Ca$^{2+}$: 1.12 \AA. Square planar 
Cu$^{2+}$ is 0.55 \AA; unfortunately Ni$^{1+}$ 
is so rare that no value is agreed, but extrapolating from other formal valences
a reasonable guess is 10-15\% larger. These differences account for the smaller volume
CCO. Of importance for NNO is $k_z$ dispersion of the $d_{z^2}$ bands,
which is larger in NNO due to hopping through the Nd $5d$ orbitals in spite of
the larger interlayer separation in NNO. The
effects of the different Madelung potentials, lowering of the O $2p$ bands
in NNO, have been discussed in several of the papers cited in Sec. I.

In the GGA+U calculations,
for the Nd ions, the Hubbard $U^{Nd}_f=8.0$ eV and the Hund's exchange coupling
$J^{Nd}_f=1.0$ eV are fixed in all calculations,
which provide the $S=\frac{3}{2}$ Nd$^{3+}$ spin state \cite{choi2020}. 
The total angular momentum ${\cal J}$=$\frac{9}{2}$ gives a correspondingly large total magnetic moment.
We varied $U^{Ni}_d$ in the range of 0--5 eV, to study correlation effects up to
the value that is reasonable for Ni ions in a conducting oxide, and
with fixed $J^{Ni}_d=0.7$ eV.
We quote values of $U^{Ni}_{d}$ (simply called $U$) that we explore.
As discussed in following sections, including $U$ on Ni is essential 
but the electronic structure
near the Fermi level is practically invariant to $U$ in the 2--5 eV range.
We have mostly focused on results for $U=4$ eV.

In {\sc wien2k},
the basis size was determined by $R_{mt}K_{max}=8$ and
the augmented-plane-wave radii $R_{mt}$ of Nd 2.50, Ni 1.95, and O 1.65, in atomic units.
The other input parameters of these calculations are the same as in our 
previous study of NdNiO$_2$ \cite{choi2020}.

\section{Energies and moments} 
The (110) AFM state we study, within a $\sqrt{2}\times\sqrt{2}$ supercell, 
has both antialigned Nd and Ni spin layers, with the structure giving no net near neighbor
Nd-Ni coupling. This AFM ordering is energetically favored over aligned Ni spins by 
95 (62) meV/f.u. at $U=0$ (4) eV, both being strongly favored over nonmagnetic Ni \cite{choi2020,botana2020}. 
A theoretical value of the energy for local but disordered Ni moments
requires separate calculations, but at maximum would be of the order of $J^{Ni}_d$.
Leonov {\it et al.}, for example, obtain first and second neighbor exchange
constants that put it in the regime of magnetic frustration
within the $J_1$-$J_2$ spin model only upon electron doping \cite{leonov2020}.
Although DFT often overestimates magnetic moments and thereby
energies in weak magnets, it is commonly accurate for larger (ionic) moments. These Ni moments are
of the order of 1$\mu_B$. 
Adding the effect of $U$ revises the energetics and increases the moments somewhat,
but not the relative stabilities of AMF and FM order.
The observed avoidance by NNO of a relatively large ordering energy is our focus.

\begin{figure}[tpb]
%\vskip 8mm
 {\resizebox{8cm}{9cm}{\includegraphics{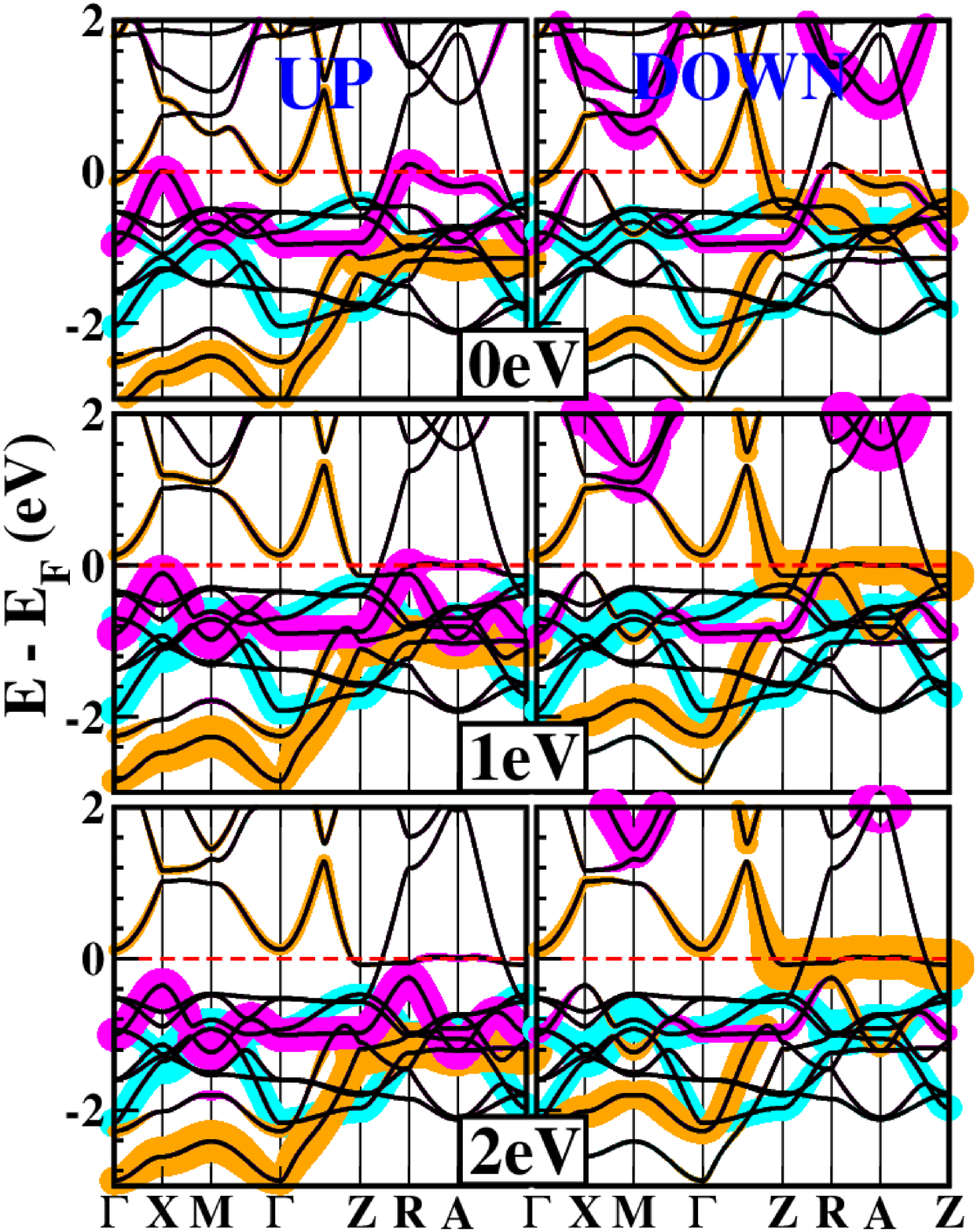}}}
\caption{Fatband depiction of the AFM band structure in GGA+U, $U$=0, 1, 2 eV:
(Left) Ni majority; (right) Ni minority; E$_F\equiv 0$.
Characters that are emphasized are Ni $d_{z^2}$ (orange), $d_{x^2-y^2}$
(magenta), and $d_{xy}$ (blue), relative to the primitive cell.   
The $d_{z^2}$ band flattens along the $Z-R-A-Z$ lines ($k_z=\pi/c$ zone face) 
already for $U$=1 eV.
Relative to the primitive cell,
symmetry points are rotated by 45$^\circ$ in the $\sqrt{2}\times\sqrt{2}$ AFM supercell, 
{\it i.e.}, $M(R)\leftrightarrow X(A)$.
}
\label{band}
\end{figure}

\section{Correlation effects and the flat band}
Correlation effects are immediate and unexpected. One thing that proceeds as expected
is that the Ni $d_{x^2-y^2}$ bands are split by the order of $U$; in cuprates this
opens a gap. In the nearest neighbor tight binding picture these bands are perfectly 
nested in the diamond-shaped region that becomes the Brillouin zone (BZ) in the AFM zone. From either a
perfect nesting perspective or a superexchange spin model, a gap is opened and AFM
order is robust.

However, including even a small value of $U$ leads to  
a dramatic change in the Ni $d_{z^2}$-derived band initially below but very near the Fermi energy $E_F$.
This band shows $k_z$ dispersion assisted by the Nd $d$ orbital, but negligible
dispersion for $k_z=\pi/c$ ($Z-R-A-Z$ lines), where the bands remain degenerate
in spite of being half filled.

Figure~\ref{band} shows in fatband form the Ni $d_{z^2}$ (orange), $d_{x^2-y^2}$
(magenta), and $d_{xy}$ (blue) 
band characters as $U$ increases over the range 0--2 eV; $d_{xz}, d_{yz}$ orbitals of both
spin are occupied. At $U$=0, all three
orbitals are active near E$_F$. However, already with the modest value $U$=1 eV,  an 
unanticipated flatness of the $d_{z^2}$ band emerges precisely at E$_F$ along the 
entire top zone face $k_z=\pi/c$ while some (but small) dispersion remains in the
$k_z$=0 plane.  This distinctive flat band remains
in place as $U$ increases to 5 eV, not shown in the figure. To emphasize:
the flat band on the upper/lower zone face is a robust feature of AFM order 
and two band behavior, with implications discussed below.

The relatively short Ni-Ni distance 1.685 \AA~along the $c$-axis induces a hopping
(assisted by Nd $d_{z^2}$ and $d_{xy}$) between the $d_{z^2}$ orbitals on nearby layers, 
leading to dispersion of 2 eV along the $\Gamma-Z$ line.
For NNO, the related effective hopping parameter is $t_{d_{z^2}}\approx 1$ eV, 
large for $k_z$ hopping across a 3.4 \AA~distance but aided by the Nd $5d$ orbital.
For the minority (higher energy) states, the Ni $d_{z^2}$ orbital is mixed with Nd $d_{xy}$ 
and somewhat with Nd $d_{x^2-y^2}$ on the $k_z=\pi/c$ plane, resulting in a 
hybridization gap of 0.4 eV at the $Z$-point.
The mixtures elsewhere, {\it viz.} at $\Gamma$, are negligible.

\begin{figure}[tbp]
%\vskip 8mm
{\resizebox{8cm}{6.0cm}{\includegraphics{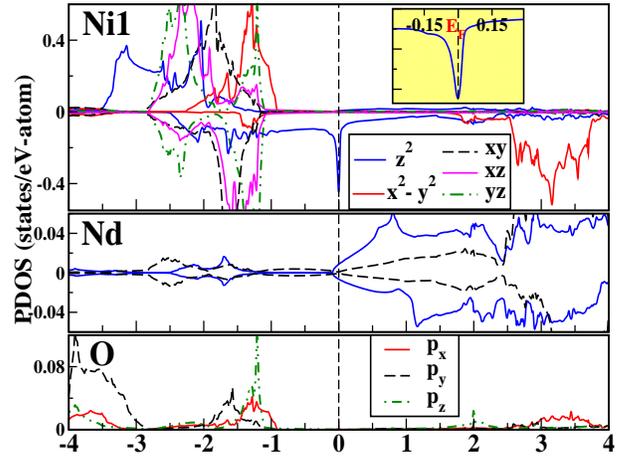}}}
\caption{
AFM orbital-projected densities of states (PDOSs)/atom of Ni $3d$, two Nd $5d$ $d_{z^2}$ and $d_{xy}$ orbitals, 
and O $2p$ orbitals, for the undistorted lattice
at $U=4$ eV, in the $3d$ band region (note the difference in scales).
Majority (minority) are plotted upward (downward).
The oxygen PDOS lies below 3 eV. Having E$_F$ pinned precisely at the sharp, 1D-like vHs
(upper panel, minority spin, shown enlarged in the Inset) 
produces the  instabilities discussed in the text.
}
\label{dos}
\end{figure}

Figure \ref{dos} shows (for the realistic metallic value $U$=4 eV)
Ni, Nd, and O orbital-projected densities of states (PDOSs) in the $d$ band 
region surrounding E$_F$. The unfilled minority Ni $d_{x^2-y^2}$ orbital
(the hole) lies at +3 eV. One remarkable feature is that the minority Ni $d_{z^2}$
band spans nearly 5 eV, crossing $E_F$ and ensuring a conducting state. All other
Ni $d$ orbitals of both spins are narrow. The other notable feature is that
this same orbital gives rise to
a flat band across the entire $k_z=\pi/c$ zone face that produces a 1D van Hove singularity 
(vHs) {\it pinned at $E_F$}, with pure $d_{z^2}$ character; 
recall, these two exotic features are robust, that is, insensitive to $U$.
The valence bands have nearly pure Ni $d_{z^2}$ character within 1 eV 
of $E_F$. Nd $5d$ character extends down to and slightly below 
E$_F$ as noted several times previously,
but does not participate in the flat band.
Around $Z$ (see along the $Z-R$ and $Z-A$ lines in Fig.~\ref{band}), the flat band 
crosses a dispersive band with mixed Nd $5d_{z^2}$, 
Ni $3d_{xz/yz}$ character at $E_F$. 

As a result of coupling through Nd states, the  
flat band acquires a small width $\sim$40 meV corresponding to an effective 
{\it in-plane} hopping $t^{eff}_{z^2}$
=20 meV for this plane.  
Based on charge decompositions and PDOSs, and the Ni $d_{z^2}$ character above E$_F$, 
the evidence suggests a formal valence closer to  
Ni$^{1.4+}$ rather than Ni$^{1+}$ as would be appropriate for an insulator. (Since
O is clearly $2-$, we expect this is a breakdown of formal valence counting as
often occurs in metals.)
In contrast to cuprates where the central role implicates the Cu $d_{x^2-y^2}$ orbital,
our results support previous indications \cite{lee2004} that the Ni $d_{z^2}$ orbital becomes a central 
player in NNO (and LNO), with the $d_{x^2-y^2}$ hole being more of a given and less of a
dynamical component.

\begin{figure}[tbp]
%\vskip 8mm
{\resizebox{6cm}{3.5cm}{\includegraphics{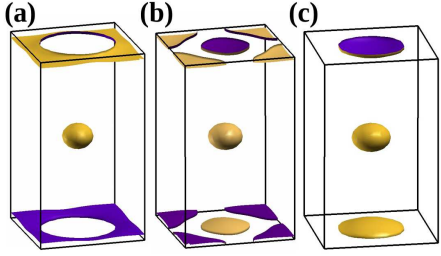}}}
\caption{
Variations of AFM Fermi surfaces by carrier dopings:
(a) 0.025 hole-doping, (b) undoped, and (c) 0.025 electron-doping per f.u. 
at $U=4.0$ eV.
Small doping levels (virtual crystal) were chosen to probe the geometry
of the (nearly) flat band.
}
\label{fs}
\end{figure}

\section{Fermi surfaces versus doping}
To probe the geometry of the (nearly) flat band, the AFM Fermi surfaces (FSs) 
for $U=4$ eV versus small (virtual crystal) doping levels are shown in 
Fig.~\ref{fs}. A $\Gamma$-centered electron sphere, a mixture of Ni $3d_{z^2}$ and 
Nd $5d_{z^2}$ characters \cite{lee2004,choi2020}, 
arises from the high velocity band and slowly changes volume upon doping. 
2D wafers appear near the $k_z=\pm\pi/c$ planes. At stoichiometry,
a round hole wafer is centered at $Z$ and a rounded-diamond electron wafer at 
the zone corner $A$.
For hole-doping, the thin hole wafer is pinched off around the $Z$-point
[Fig.~\ref{fs}(a)], and the electron wafers connect.
Small electron doping, Fig.~\ref{fs}(c), only electron buttons appear around $Z$.

The flat and thin wafer and button FSs at stoichiometry have a characteristic thickness 
$2K^F_z\leq$0.06$\frac{\pi}{c}$. In addition to the flatness, the very small
velocities will lead to strong nesting, at $K^F_z$ and $\frac{2\pi}{c}-2K^F_z$.
The related susceptibility will encourage spin density, charge density, and lattice
fluctuations, and possibly instabilities at $\sim 32c$ and $\sim 2c$ respectively.
The long wavelength instability would be challenging to study numerically, the short
wavelength one corresponds to a dimerization of successive NiO$_2$ layers -- a
Peierls instability of the $d_{z^2}$ band and its associated vHs peak.

\begin{figure}[tbp]
%\vskip 8mm
{\resizebox{6cm}{8cm}{\includegraphics{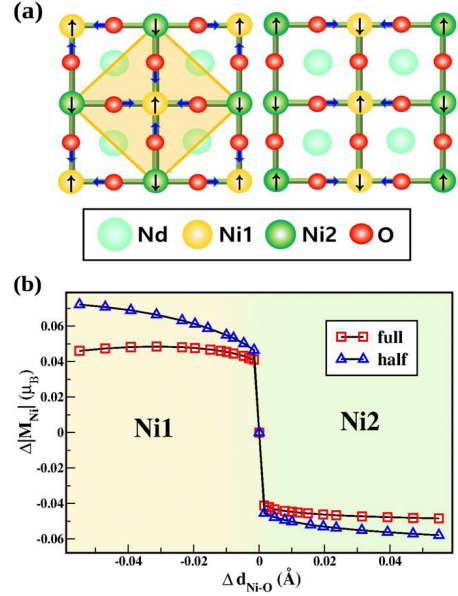}}}
\caption{%{\bf Evidence of symmetry breaking.}
(a) Sketch of of the NiO$_2$ layer with (left) oxygen full-breathing mode 
and (right) half-breathing mode along (100) direction,
showing $2\times 2$ primitive cells.
The full- and half-breathing modes allow for checkerboard and stripe 
charge and spin orders, respectively.
The arrows on the Ni ions indicate spins of Ni ions, whereas the 
blue (thick) arrows denote the displacement of oxygen ions.
(b) Changes of the magnitude of Ni magnetic moments $\Delta M$, 
versus changes in the Ni-O bond length $\Delta d_{Ni-O}$
for both the full- and half-breathing modes at $U=4$ eV.
This `spin  disproportionation' instability occurs for the undistorted lattice.
There is an accompanying charge disproportionation (see text).
}
\label{sym}
\end{figure}

\section{Magnetic Instability}
A flat band generated DOS peak at E$_F$ in an AFM metal causes a
possible ``meta-Stoner'' instability of the AFM order
(Stoner instability being with respect to the 
already spin-polarized state).  
To break the underlying spin symmetry in this already magnetically-ordered AFM state,
the spin symmetry must be broken. We have studied this magnetic instability
together with lattice instability by imposing breathing  and half-breathing 
types of oxygen distortions,
see Fig.~\ref{sym}(a), which destroy the equivalence of the Ni ions. 
For the former, the unit cell is not 
enlarged; for the latter, the cell is doubled, with Ni ions of each spin
direction being squeezed, or not, by the half-breathing distortion. 
The range of displacements is shown in Fig.~\ref{sym}(b).

For the full- and half-breathing modes,
the changes in moments of the two Ni sites, Ni1 and Ni2,  
are displayed in Fig.~\ref{sym}(b). 
These frozen oxygen phonon modes lead to an unexpected change in the Ni magnet moments
even for a very small displacement, indicating the magnetic instability is there even for
the symmetric lattice.
Tiny mode amplitudes (displacements down to 0.001 \AA~extrapolating to zero; 
allowing for symmetry breaking is what is important) 
produce a `spin disproportionation' of the AFM ions
of the order of $\pm$0.05 $\mu_B$.  
This behavior signals that the AFM state
is unstable to a first order transition to a {\it ferri}magnetic state, 
which reflects the partially itinerant character of the Ni moments. 
The mechanism is the 
vHs generated meta-Stoner instability, with feedback within the self-consistency 
loop (necessitated by the vHs) providing the first order nature; actual lattice
distortion is unnecessary. The magnetic symmetry breaking is accompanied by a 
Ni1-to-Ni2 charge disproportionation up to $\sim\pm$0.05$e$ 
(from atomic sphere charges). 
Being accompanied by the charge disproportionation as well,
these results suggest there would be substantial magnetic and charge fluctuations
in the AFM state, contributing to its evident instability.

\begin{figure}[tbp]
%\vskip 8mm
{\resizebox{8cm}{6cm}{\includegraphics{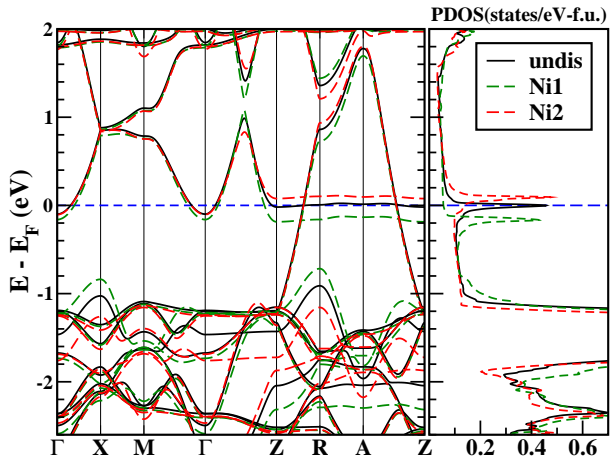}}}
\caption{
AFM band structures and Ni atom-PDOSs for $U=4$ eV.
The (red and green) dashed lines indicate the bands split by the full-breathing mode,
with O displaced by 0.03 \AA. 
The energy splitting of each band per unit displacement is 5 eV/\AA. 
The half-breathing mode (not shown) shows similar band splitting.
The peak-splitting is the source of the Peierls (meta-Stoner) charge (spin) instability.
}
\label{bs_dist}
\end{figure}

\begin{figure}[!ht]
%\vskip 8mm
{\resizebox{8cm}{6cm}{\includegraphics{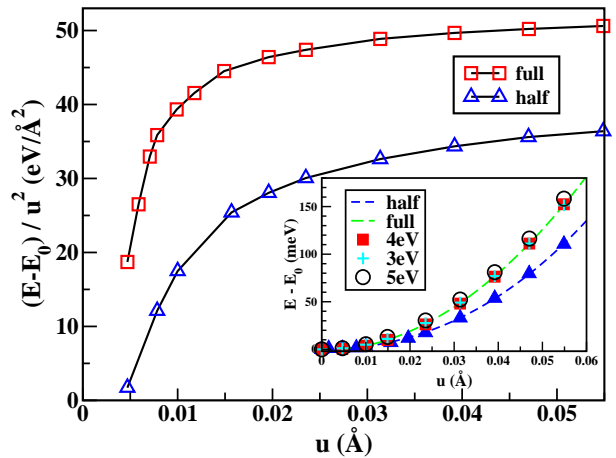}}}
\caption{
Energy distortions for the frozen-in oxygen full- and half-breathing modes at $U=4$ eV.
As shown in the Inset, these follow a quadratic plus lowest-order anharmonic term,
indicated by the green (blue) dashed lines,
and are insensitive to strength of $U$ in the range of 3--5 eV.
}
\label{ph}
\end{figure}

\section{Lattice Instability} 
The effect of the breathing mode on the band structure is
displayed in Fig.~\ref{bs_dist}, quantifying the splitting of the vHs peak. 
The splitting of the peak at $E_F$ due to the frozen phonon modes
leads to a characteristic deformation potential 
(splitting of each band per unit displacement) of
%  ${\cal D}=3.75$ eV/\AA~ on the $k_z=c/\pi$.\cite{LP2004}
 ${\cal D}$=150 meV/0.03 \AA=5 eV/\AA~ on the $k_z=\pm\pi/c$ planes \cite{LP2004}.
This value places AFM NNO in the strong electron-phonon coupling regime.
Due to the vHs giving strong structure in the density of states
$N(E)$, this splitting of the vHs peak varies in the range of
3.1 to 7.3 eV/\AA~ for displacements in the range 0.01 to 0.047 \AA.
This variation leads to strong effects on the electronic and magnetic properties,
as discussed below.

To explore the Peierls lattice instability, we plot the energy difference
divided by squared amplitude $K(u)$=$\Delta E(u)$/$u^2$ in Fig.~\ref{ph};
this is the lattice stiffness (effective force constant) $K$. 
For a harmonic phonon $K(u)$ would be a constant, with deviation indicating
anharmonicity. 
The energy change with oxygen displacement $u$, $\Delta E(u)=E(u)-E(0)$,
can be fit very well by $A_2u^2+A_3u^3$.
The obtained coefficients are \\
~~~ $A_2=47.5$ eV/\AA$^2$, $A_3=61.2$ eV/\AA$^3$ (breathing);\\
~~~ $A_2=29.4$ eV/\AA$^2$, $A_3=136.7$ eV/\AA$^3$ (half-breathing).\\
As shown in the inset of Fig. 6,
these values are insensitive to the strength of $U$ in the range of 3--5 eV.
The harmonic terms provide the frequency of each breathing mode (see below),
whereas the $A_3u^3$ terms quantify anharmonicity.

The calculated $K(u)$ is anomalous in two ways. For typical
oxygen vibrational amplitudes 0.05-0.10 \AA, it would be practically constant;
high frequency O modes typically do not show significant anharmonicity. 
Figure~\ref{ph} however shows that for both breathing and half-breathing
modes, $K(u)$ is only beginning to approach a steady value somewhere above 
0.05 \AA. 
Secondly, $K(u)$ drops precipitously below 0.02 \AA~and extrapolates to 
{\it negative} values below 0.005 \AA, {\it viz.} giving the classic unstable
Mexican hat potential surface. Modern linear response lattice dynamics 
codes that incorporate infinitesimal distortions would calculate imaginary
frequencies in both cases, the signature of lattice instabilities for both
types of modes. This behavior is revealing the response of the electronic
vHs to (infinitesimal) perturbation, analogous to the magnetic instability
discussed above.

If calculated as often done
on a mesh of $u$ such as 0.01, 0.02, 0.03 \AA~ etc., 
the unstable region would
not be sampled carefully, but anharmonicity [non-constancy of $K(u)$] would be evident;
high energy modes around $\omega=80$ meV and 60 meV for the full- and 
half-breathing modes respectively would be obtained, similar to our values.
Compared with the values for these modes in CaCuO$_2$ \cite{cohen2007} 
and in La$_2$CuO$_4$ \cite{lco_exp2005},
these frequencies are smaller by 10\%-25\% respectively. 
Both of these compounds are insulators with no important anharmonicity
being reported.

The lattice and magnetic instabilities reflect the sharp and narrow vHs 
that negates normal charge response and provides a platform that is
extremely sensitive to any symmetry breaking that splits the vHs peak. 
The most anomalous response is confined to small distortions and energies 
(and thus low temperature), as is clear from Fig.~\ref{ph}.
For this Peierls instability, a fully quantum treatment
would require self-consistent treatment of (1) anharmonicity, 
(2) non-adiabatic electron-lattice coupling around the vHs point
(where heavy charge carriers are coupled to phonons of similar, or even 
higher, energy), and (3) behavior of coupled quasiparticle self-energies,
which is far beyond the scope of this paper.  

A fully quantum treatment would include also the effects of oxygen zero point 
fluctuations (ZPFs) on the system and its response. 
For frequencies of 60-80 meV, the oxygen zero point amplitude  
   %=\sqrt{\frac{\hbar}{2M\omega}}$
$\langle u_0\rangle\approx$ 0.04-0.05 \AA. Within this distance of the oxygen site, the potential
becomes strongly anharmonic and finally negative. 
To what extent ZPFs could temper the static 
lattice, spin, and charge instabilities might be challenging to determine, 
because ZPF effects are subtle: the fully quantum metal has zero resistivity at T=0, because
ZPFs do not scatter carriers, whereas the simple picture of averaging over
displaced atoms would (incorrectly) suggest otherwise.  

An important dynamical lattice consequence for this
AFM system is that the vHs will be renormalized at low temperature -- 
narrowed instead of the naive expectation of broadening \cite{quan2019} by 
electron-phonon interaction -- and the issue
of the behavior of the non-adiabatic coupled electron-phonon excitations is not settled.   
Just how the massive flat band 
carriers would deal with all of these instabilities would be a
challenging theoretical problem, but the experimental outcome is that the underlying 
AFM order we have modeled gives way to a disordered moment metal.

\section{Discussion: avoidance of the AFM phase}
We have explored the experimentally inaccessible AFM ordered state of NdNiO$_2$ with 
correlated DFT methods. This is the state whose calculated energy indicates it 
should be the ground state versus nonmagnetic and competing simple magnetic states, 
and presumably the disordered moment phase as well, seemingly a severe problem for
theory. 
However, we find as well that this state is multiply unstable: lattice,
spin, or charge fluctuations -- neglected in this static lattice, fixed spin
DFT modeling -- so further studies would be necessary to progress to 
a determination and understanding of the ground state. 
Experiments indicate a conducting state 
without magnetic order, but likely with fluctuating moments
(Ni and Nd) that make it very challenging to model from a DFT viewpoint.

Specifically, we have shown that in the (unphysical) AFM ordered phase
a Ni $d_{z^2}$ flat band pinned at the Fermi level
arises on the entire $k_z=\pm\pi/c$ zone faces,
giving rise to a 1D-like van Hove singularity that supports
spin, charge, and lattice (both full- and half-breathing mode) instabilities of the
ideal infinite-layer lattice.
Due to the narrowness of the vHs, these coupled order parameters may be 
affected by quantum
zero point oxygen motion and non-adiabatic electron-lattice coupling in the vHs peak, 
finally frustrating these types of symmetry breaking. These observations account
qualitatively for the stabilization of the observed symmetric (conducting,
no spin order) phase.

Yet, the AFM ordered state, calculated to be most stable for the static lattice,
remains inaccessible by experiment, hence evidently is higher in 
{\it free energy} than the conducting spin-disordered state. We propose that
as temperature is lowered, NdNiO$_2$ approaches the AFM ordered phase but
encounters its incipient instabilities with strong spin, charge, and lattice fluctuations
that inhibit spin order and lattice distortion, thereby stabilizing the 
symmetric lattice. 
The result is to remain spin-disordered but with AFM correlations, {\it i.e.},
incipient AFM order, that will reduce the spin
free energy. This correlated spin liquid phase,
the Sachdev-Read MQPM phase that has been proposed in another disordered
moment superconductor\cite{stpi2019},
provides the platform for the superconductivity that appears upon hole doping.  
None of these complications occur in CaCuO$_2$.

\section{Acknowledgments}
We acknowledge Sooran Kim for a critical reading of the manuscript.
M.Y.C. and K.W.L were supported by National Research Foundation of Korea
Grant No. NRF-2019R1A2C1009588. 
W.E.P. was supported by NSF Grant No. DMR 1607139.


\begin{thebibliography}{10}

%%%% model theory %%%%

\bibitem{c.wu2019}L.-H. Hu and C Wu,
  Two-band model for magnetism and superconductivity in nickelates,
  Phys. Rev. Research {\bf 1}, 032046 (2019).

\bibitem{leonov2020}I. Leonov, S. L. Skornyakov, and S. Y. Savrasov,
 Lifshitz transition and frustration of magnetic moments in infinite-layer
  NdNiO$_2$ upon hole doping,
 Phys. Rev. B {\bf 101}, 241108(R) (2020).

\bibitem{mjhan2020}S. Ryee, H. Yoon, T. J. Kim, M. Y. Jeong, and M. J. Han,
Induced magnetic two-dimensionality by hole doping in the superconducting infinite-layer nickelate
Nd$_{1-x}$Sr$_x$NiO$_2$,
Phys. Rev. B {\bf 101}, 064513 (2020).

\bibitem{wu2020}X. Wu, D. D. Sante, T. Schwemmer, W. Hanke, H. Y. Hwang, 
  S. Raghu, and R. Thomale,
 Robust $d_{x^2-y^2}$-wave superconductivity of infinite-layer nickelates,
 Phys. Rev. B {\bf 101}, 060504(R) (2020).


\bibitem{werner2020}P. Werner and S. Hoshino,
Nickelate superconductors: Multiorbital nature and spin freezing,
Phys. Rev. B {\bf 101}, 041104(R) (2020).

\bibitem{zhang2020}G.-M. Zhang, Y.-F. Yang, and F.-C. Zhang,
Self-doped Mott insulator for parent compounds of nickelate superconductors,
Phys. Rev. B {\bf 101}, 020501(R) (2020).

\bibitem{lechermann2020} F. Lechermann,
 Late transition metal oxides with infinite-layer structure: Nickelates versus cuprates,
Phys. Rev. B {\bf 101}, 081110(R) (2020).


\bibitem{zhang-jin2020}H. Zhang, L. Jin, S. Wang, B. Xi, X. Shi, F. Ye, and J.-W. Mei,
Effective Hamiltonian for nickelate oxides Nd$_{1-x}$Sr$_x$NiO$_2$,
Phys. Rev. Research {\bf 2}, 013214 (2020).


\bibitem{kuroki2019}H. Sakakibara, H. Usui, K. Suzuki, T. Kotani, H. Aoki, and K. Kuroki,
 Model construction and a possibility of cuprate-like pairing in a new $d^9$ 
 nickelate superconductor (Nd,Sr)NiO$_2$, 
 Phys. Rev. Lett. {\bf 125}, 077003 (2020).


\bibitem{gu2019}Y. Gu, S. Zhu, X. Wang, J. Hu, and H. Chen,
 A substantial hybridization between correlated Ni-d orbital and itinerant electrons 
 in infinite-layer nickelates, 
 Commun. Phys. {\bf 3}, 84 (2020).

\bibitem{chang2019}J. Chang, J. Zhao, and Y. Ding,
 Hund-Heisenberg model in superconducting infinite-layer nickelates,
 arXiv:1911.12731.

\bibitem{berciu2020}M. Jiang, M. Berciu, and G. A. Sawatzky,
 Critical nature of the Ni spin state in doped NdNiO$_2$,
  Phys. Rev. Lett. {\bf 124}, 207004 (2020).

\bibitem{zhao2019}T. Zhou, Y. Gao, and Z. D. Wang,
 Spin excitations in nickelate superconductors,
 Sci. China-Phys. Mech. Astron. {\bf 63}, 287412 (2020).

\bibitem{karp2020}J. Karp, A. S. Botana, M. R. Norman, H. Park,
 M. Zingl, and A. Millis,
 Many-body Electronic Structure of NdNiO$_2$ and CaCuO$_2$,
 Phys. Rev. X {\bf 10}, 021061 (2020).

%%%% DFT, DMFT papers %%%%

\bibitem{anisimov1999}V. I. Anisimov, D. Bukhvalov, and T. M. Rice,
  Electronic structure of possible nickelate analogs to the cuprates,
 Phys. Rev. B {\bf 59}, 7901 (1999).

\bibitem{lee2004}K.-W. Lee and W. E. Pickett,
 Infinite-layer nickelate LaNiO$_2$: Ni$^{1+}$ is not Cu$^{2+}$,
 Phys. Rev. B {\bf 70}, 165109 (2004).

\bibitem{choi2020} M.-Y. Choi, K.-W. Lee, and W. E. Pickett,
 Role of 4f states in infinite-layer NdNiO$_2$,
 Phys. Rev. B {\bf 101}, 020503(R) (2020).

\bibitem{botana2020} A. S. Botana and M. R. Norman,
 Similarities and Differences between LaNiO$_2$ and CaCuO$_2$ and
  Implications for Superconductivity,
  Phy. Rev. X {\bf 10}, 011024 (2020).

\bibitem{nomura2019} Y. Nomura, M. Hirayama, T. Tadano, Y. Yoshimoto, K. Nakamura, and R. Arita,
Formation of a two-dimensional single-component correlated electron system
and band engineering in the nickelate superconductor NdNiO$_2$,
Phys. Rev. B {\bf 100}, 205138 (2019).

\bibitem{jiang2019} P. Jiang, L. Si, Z. Liao, and Z. Zhong,
Electronic structure of rare-earth infinite-layer RNiO$_2$(R=La,Nd),
Phys. Rev. B {\bf 100}, 201106(R) (2019).


\bibitem{ZLiu2019}Z. Liu, Z. Ren, W. Zhu, Z. F. Wang, and J. Yang,
 Electronic and Magnetic Structure of Infinite-layer NdNiO$_2$: 
  trace of antiferromagnetic metal, 
  npj Quantum Mater. {\bf 5}, 31 (2020).
 
\bibitem{geisler2020}B. Geisler and R. Pentcheva,
 Fundamental difference in the electronic reconstruction of 
 infinite-layer versus perovskite neodymium nickelate films on SrTiO$_3$(001), 
 Phys. Rev. B {\bf 102}, 020502(R) (2020).

%%%% end of DFT, DMFT papers
  

%%%% exptl papers on NNO2 %%%%%

\bibitem{H.Hwang2019}D. Li, K. Lee, B. Y. Wang, M. Osada, S. Crossley,
   H. R. Lee, Y. Cui, Y. Hikita, and H. Y. Hwang,
  Superconductivity in an infinite-layer nickelate,
 Nature (London) {\bf 572}, 624 (2019).

%recent published papers
\bibitem{hepting2020} M. Hepting,
  D. Li, C. J. Jia, H. Lu, E. Paris, 
 Y. Tseng, X. Feng, M. Osada, E. Been, Y. Hikita, Y.-D. Chuang, Z. Hussain, 
 K. J. Zhou, A. Nag, M. Garcia-Fernandez, M. Rossi, H. Y. Huang,
 D. J. Huang, Z. X. Shen, T. Schmitt, H. Y. Hwang, B. Moritz, J. Zaanen, 
 T. P. Devereaux and W. S. Lee,
  Electronic structure of the parent compound of superconducting 
  infinite-layer nickelates, 
 Nat. Mater. {\bf 19}, 381 (2020).

\bibitem{QLi2019}Q. Li, C. He, J. Si, X. Zhu, Y. Zhang, and H.-H. Wen,
 Absence of superconductivity in bulk 
Nd$_{1-x}$Sr$_x$NiO$_2$,
 Commun. Mater. {\bf 1}, 16 (2020).  %arXiv:1911.02420.

\bibitem{fu2020} Y. Fu,
 Y. Fu, A. Wang, H. Cheng, S. Pei, X. Zhou, J. Chen, 
 S. Wang, R. Zhao, W. Jiang, C. Liu, M. Huang, 
 X. Wang, Y. Zhao, D. Yu, F. Ye, S. Wang, and J.-W. Mei,
 Core-level x-ray photoemission and Raman spectroscopy studies on 
  electronic structures in Mott-Hubbard type nickelate oxide NdNiO$_2$,
 arXiv:1911.03177.

\bibitem{cco110}M. Azuma, Z. Hiroi, M. Takano, Y. Bando, and Y. Tekeda,
  Superconductivity at 110 K in the infinite-layer compound
   (Sr$_{1-x}$Ca$_{x}$)$_{1-y}$CuO$_2$,
  Nature (London) {\bf 356}, 775 (1992).

%%% end of NNO2 exptl papers %%%


%EXP of LNO
\bibitem{crespin} M. Crespin, P. Levitz, and L. Gatineau,
 Reduced forms of LaNiO$_3$ perovskite. Part 1. Evidence for new phases: La$_2$Ni$_2$O$_5$ and LaNiO$_2$,
 J. Chem. Soc., Faraday Trans. 2 {\bf 79}, 1181 (1983).

\bibitem{hayward99} M. A. Hayward, M. A. Green, M. J. Rosseinsky, and J. Sloan,
 Sodium Hydride as a Powerful Reducing Agent for Topotactic Oxide 
  De-intercalation:~Synthesis and Characterization 
 of the Nickel(I) Oxide LaNiO$_2$,
 J. Am. Chem. Soc. {\bf 121}, 8843 (1999).


\bibitem{kaneko09} D. Kaneko, K. Yamagishi, A. Tsukada, T. Manabe, and M. Naito,
 Synthesis of infinite-layer LaNiO$_2$ films by metal organic decomposition,
  Physica C: Superconductivity {\bf 469}, 936 (2009).

\bibitem{ikeda13} A. Ikeda, T. Manabe, and M. Naito,
 Improved conductivity of infinite-layer LaNiO$_2$ thin films by metal organic decomposition,
 Physica C: Superconductivity {\bf 495}, 134 (2013).




\bibitem{sachdev}S. Sachdev and N. Read,
 Metallic spin glasses,
 J. Phys.: Condens. Matt. {\bf 8}, 9723 (1996).

\bibitem{werner1}S. Hoshino and P. Werner,
 Superconductivity from Emerging Magnetic Moments,
  Phys. Rev. Lett. {\bf 115}, 247001 (2015).

\bibitem{werner2}P. Werner, S. Hoshino, and H. Shinaoka,
Spin-freezing perspective on cuprates,
  Phys. Rev. B {\bf 94}, 245134 (2016).

\bibitem{stpi2019}S.-T. Pi, S. Y. Savrasov, and W. E. Pickett,
 Pressure-tuned frustration of Magnetic Coupling in Elemental Europium,
 Phys. Rev. Lett. {\bf 122}, 057201 (2019).


\bibitem{pardo2011}V. Pardo and W. E. Pickett,
  Metal-insulator transition in layered nickelates La$_3$Ni$_2$O$_{8-\delta}$
  ($\delta=0.0, 0.5, 1$),
  Phys. Rev. B {\bf 83}, 245128 (2011). 


\bibitem{botana2016}A. S. Botana, V. Pardo, W. E. Pickett, and M. R. Norman,
  Charge ordering in Ni$^{1+}$/Ni$^{2+}$ nickelates: La$_4$Ni$_3$O$_8$ and La$_3$Ni$_2$O$_6$,
  Phys. Rev. B {\bf 94}, 081105(R) (2016).


\bibitem{mimicking} J. Zhang, A. S. Botana, J. W. Freeland, D. Phelan, H. Zheng, 
  V. Pardo, M. R. Norman, and J. F. Mitchell,
  Mimicking cuprates: large orbital polarization in a metallic square-planar nickelate,
  Nat. Phys. {\bf 13}, 864 (2017).


\bibitem{nickelates}A. S. Botana, V. Pardo, and M. R. Norman,
  Electron doped layered nickelates: spanning the phase diagram of the cuprates,
  Phys. Rev. Materials {\bf 1}, 021801 (2017).

\bibitem{La438} J. Zhang
   D. M. Pajerowski, A. S. Botana, Hong Zheng, L. Harriger, J. Rodriguez-Rivera, 
   J. P. C. Ruff, N. J. Schreiber, B. Wang, Yu-Sheng Chen, W. C. Chen, M. R. Norman, 
   S. Rosenkranz, J. F. Mitchell, D. Phelan
 Spin stripe order in a square planar trilayer nickelate,
  Phys. Rev. Lett. {\bf 122}, 247201 (2019).

%----------------------------------

\bibitem{gga} J. P. Perdew, K. Burke, and M. Ernzerhof,
Generalized gradient approximation made simple,
 Phys. Rev. Lett. {\bf 77}, 3865 (1996).

\bibitem{ylvisaker2010} E. R. Ylvisaker, K. Koepernik, and W. E. Pickett,
  Anisotropy and magnetism in the LSDA+U method,
 Phys. Rev. B {\bf 79}, 035103 (2009).

\bibitem{wien2k} K. Schwarz and P. Blaha,
 Solid state calculations using WIEN2k,
 Comput. Mater. Sci. {\bf 28}, 259 (2003).


\bibitem{Siegrist1988}T. Siegrist, S. M. Zahurak, D. W. Murphy, and R. S. Roth,
 The parent structure of the layered high-temperature superconductors, 
  Nature {\bf 334}, 231–232 (1988).


%B-doped diamond
\bibitem{LP2004} K.-W. Lee and W. E. Pickett,
 Superconductivity in Boron-Doped Diamond,
 Phys. Rev. Lett. {\bf 93}, 237003 (2004).


%%
%CaCuO2 breathing mode
\bibitem{cohen2007} P. Zhang, S. G. Louie, and M. L. Cohen,
Electron-Phonon Renormalization in Cuprate Superconductors,
Phys. Rev. Lett. {\bf 98}, 067005 (2007).

\bibitem{lco_exp2005} T. Fukuda, J. Mizuki, K. Ikeuchi, K. Yamada, 
 A. Q. R. Baron, and S. Tsutsui,
Doping dependence of softening in the bond-stretching phonon mode 
of La$_{2-x}$Sr$_x$CuO$_4$ ($0\le x\le0.29)$,
Phys. Rev. B {\bf 71}, 060501(R) (2005).

%%
\bibitem{quan2019} Y. Quan, S. Ghosh, and W. E. Pickett,
 Compressed Hydrides as Metallic Hydrogen Superconductors,  
  Phys. Rev. B {\bf 100}, 184505 (2019).


\end{thebibliography}
\end{document}